\documentclass[a4paper,10pt,twoside]{just}

\usepackage{multicol}
\usepackage{graphicx}
\usepackage{booktabs}
\usepackage{amssymb,bm,mathrsfs,bbm,amscd}
\usepackage[tbtags]{amsmath}
\usepackage{lastpage}

\begin{document}

\fancyhead[c]{\small  10th International Workshop on $e^+e^-$ collisions from $\phi$ to $\psi$ (PhiPsi15)}
 \fancyfoot[C]{\small PhiPsi15-\thepage}

\footnotetext[0]{Received ?? Nov. 2015}

\title{Search for the decays $\eta^{\prime} \to e^+e^-$  and $\eta \to e^+e^-$ at the VEPP-2000 $e^+e^-$ collider
\thanks{RFBR grant 15-02-03391-a, RSF project N 14-50-00080.}}

\author{%
      L.V.Kardapoltsev$^{1,2;1)}$\email{l.v.kardapoltsev@inp.nsk.su},
\\(for the SND Collaboration)
}
\maketitle

\address{%
$^1$ Budker  Institute  of  Nuclear  Physics,  SB RAS,  Novosibirsk,  630090,  Russia\\
$^2$ Novosibirsk  State  University,  Novosibirsk,  630090,  Russia\\

}

\begin{abstract}
A search for the rare decay $\eta^{\prime} \to e^+e^-$ 
has been performed with the SND detector at the VEPP-2000 $e^+e^-$ collider. 
The inverse reaction
$e^+e^- \to \eta^{\prime}$ and $\eta^{\prime}$ five decay chains have been 
used for this search. The upper limit  
$\Gamma_{\eta^{\prime} \to e^+e^-}<0.002$ eV at the 90\% confidence level 
has been set. A sensitivity of SND in a search for $\eta \to e^+e^-$ decay has 
been studied. For this perpose we have analyzed a data sample with an 
integrated luminosity of 108 nb$^{-1}$ collected in the center-of-mass energy
range 520-580 MeV. There are no background events for the reaction  
$e^+e^- \to \eta$ with decay $\eta \to \pi^0\pi^0\pi^0$ have been found.
In the absence of background, a sensitivity to $B(\eta \to e^+e^-)$ of 
10$^{-6}$ can be reached during two weeks of VEPP-2000 operation. 
\end{abstract}



\begin{multicols}{2}

\section{Introduction}
Decays of pseudoscalar mesons to the pair of leptons
$P\to l^+l^-$
are rare. In the Standard Model (SM)
these decays proceed through the two-photon intermediate state as shown in Fig.~\ref{etadec} and therefore are 
suppressed as $\alpha^2$ relative to the $P\to \gamma\gamma$ decays, where $\alpha$
is the fine structure constant. An additional suppression of $(m_l/m_P)^2$ arises from the 
approximate helicity conservation, where $m_l$ and $m_P$ are
the lepton and meson masses, respectively. So, due to the low probability
such decays are sensitive to possible contribution of new physics beyond the SM~\cite{NPh1, NPh2}.

\begin{center}
\includegraphics[width=6cm]{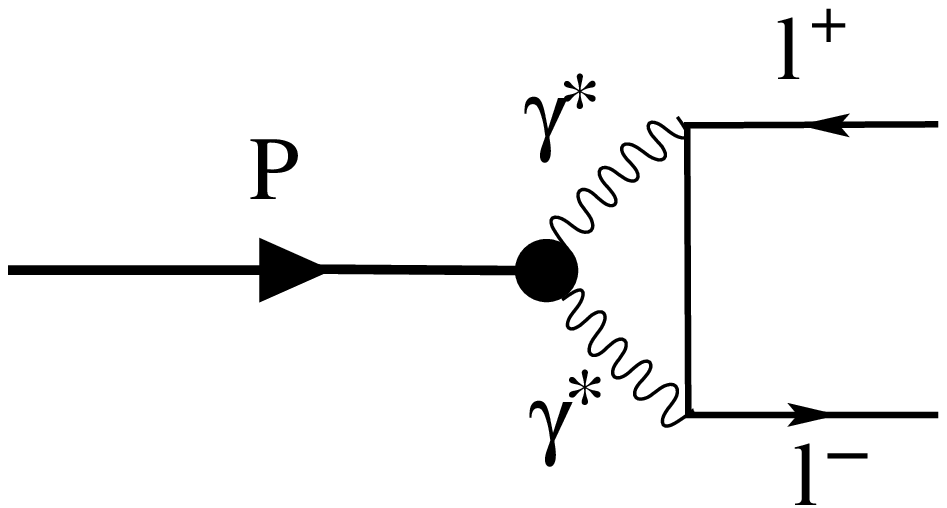}
\figcaption{\label{etadec}   Leading order QED contribution driving $P\to l^+l^-$ decays. }
\end{center}

The ranges of predictions for
the $P\to l^+l^-$
branching fractions obtained in different
form-factors models~\cite{etamod1, etamod2} are listed in Table~\ref{tab1}. For comparison, the last
column of Table~\ref{tab1} contains the current experimental values of the branching fractions.
The value of $B(\pi^0\to e^+e^-)$ differs from the theoretical prediction by about
three standard deviations. This can have different explanations connected with theoretical 
uncertainty~\cite{pi0culc} or/and with new physics contributions~\cite{NPh1, NPh2}.

This paper is devoted to the recent search for the $\eta^{\prime}\to e^+e^-$ 
decay~\cite{etapr-snd} with the SND detector~\cite{SND1,SND2}
at the VEPP-2000 $e^+e^-$ collider~\cite{vepp2000}, 
in which the inverse reaction $e^+e^- \to\eta^{\prime}$ 
is used. Also we consider 
the recent study of SND sensitivity in a search for the $\eta\to e^+e^-$ 
decay with the use of the same technique~\cite{eta-snd}.
\begin{center}
\tabcaption{ \label{tab1}  The theoretical predictions and experimental values for
the $P\to l^+l^-$ branching fractions}
\footnotesize
\begin{tabular*}{80mm}{c@{\extracolsep{\fill}}cc}
\toprule $B(P\to l^+l^-)$   & Theory  & Experiment \\
\hline
$B(\pi^0\to e^+e^-)\times 10^8$ & 6.23---6.38 & 7.49 $\pm$ 0.38~\cite{ktev}  \\
$B(\eta\to e^+e^-)\times 10^9$ & 4.60---5.24 & $<$2300~\cite{hades}  \\
$B(\eta\to \mu^+\mu^-)\times 10^7$ & 4.64---5.12 & 5.8$\pm$ 0.8~\cite{pdg}  \\
$B(\eta^{\prime}\to e^+e^-)\times 10^{10}$ & 1.15---1.86 & $<$56~\cite{etapr-snd, etapr-cmd}  \\
$B(\eta^{\prime}\to \mu^+\mu^-)\times 10^7$ & 1.14---1.36 & ---  \\
\bottomrule
\end{tabular*}
\end{center}

\section{SND detector}
The detail description of the SND detector can be found in Refs.~\cite{SND1,SND2}.
It is a nonmagnetic detector, the main part of which is a three-layer
spherical electromagnetic calorimeter based on NaI(Tl) crystals.
The solid angle covered by the calorimeter is 90\% of $4\pi$. Its energy
resolution for photons is $\sigma_E/E=4.2\%/\sqrt[4]{E({\rm GeV})}$, and
the angular resolution is about $1.5^\circ$. The directions of charged
particles are measured by a tracking system, which consists of a 9-layer
drift chamber and a proportional chamber with readout from cathode strips.
The tracking system covers a solid angle of 94\% of 4$\pi$. The
calorimeter is surrounded by a muon system, which is used, in particular,
for cosmic-background suppression.

\section{Search for $\eta^{\prime}\to e^+e^-$ decay}
For search for the decay $\eta^{\prime}\to e^+e^-$ data 
with an integrated luminosity of
about 2.9 pb$^{-1}$ are used. They were accumulated in 2013 at the c.m. energy
close to $m_{\eta^{\prime}}c^2=957.78\pm0.06$ MeV~\cite{pdg}.
During the data taking period the beam energy was monitored with an
absolute accuracy of about 60 keV by
the Back-scattering-laser-light system~\cite{emes}.
As the collider energy spread (FWHM = 0.590 MeV) is significantly larger
than the $\eta^{\prime}$ width $\Gamma_{\eta^{\prime}}=
(0.198\pm0.009)$ MeV~\cite{pdg}, the resulting cross section is proportional
to the electronic width
\begin{equation}
\sigma_{\rm vis}({\rm nb})=
(6.38\pm0.23)\Gamma_{\eta^{\prime}\to e^+e^-}({\rm eV}). 
\label{xsvis}
\end{equation}
It should be noted that the radiative corrections and the energy spread
lead to a reduction of the cross section compared to the Born one  by a factor of four.

The search for the process $e^+e^-\to\eta^{\prime}$ is performed in 
five decay chains: $\eta^{\prime} \to \eta\pi^+\pi^-$ with the $\eta$ decays
to $\gamma\gamma$ and $3\pi^0$, and $\eta^{\prime} \to \eta\pi^0\pi^0$
with the $\eta$ decays to $\pi^+\pi^-\pi^0$, $\gamma\gamma$ and 
$3\pi^0$.

Detail description of selection criteria for all decay chains can be found
in Ref.~\cite{etapr-snd}.
Only main selection parameters will be discussed in this paper.

\subsection{Decay chain $\eta^{\prime}\to\pi^+\pi^-\eta$, $\eta\to \gamma\gamma$}

\begin{center}
\includegraphics[width=6cm]{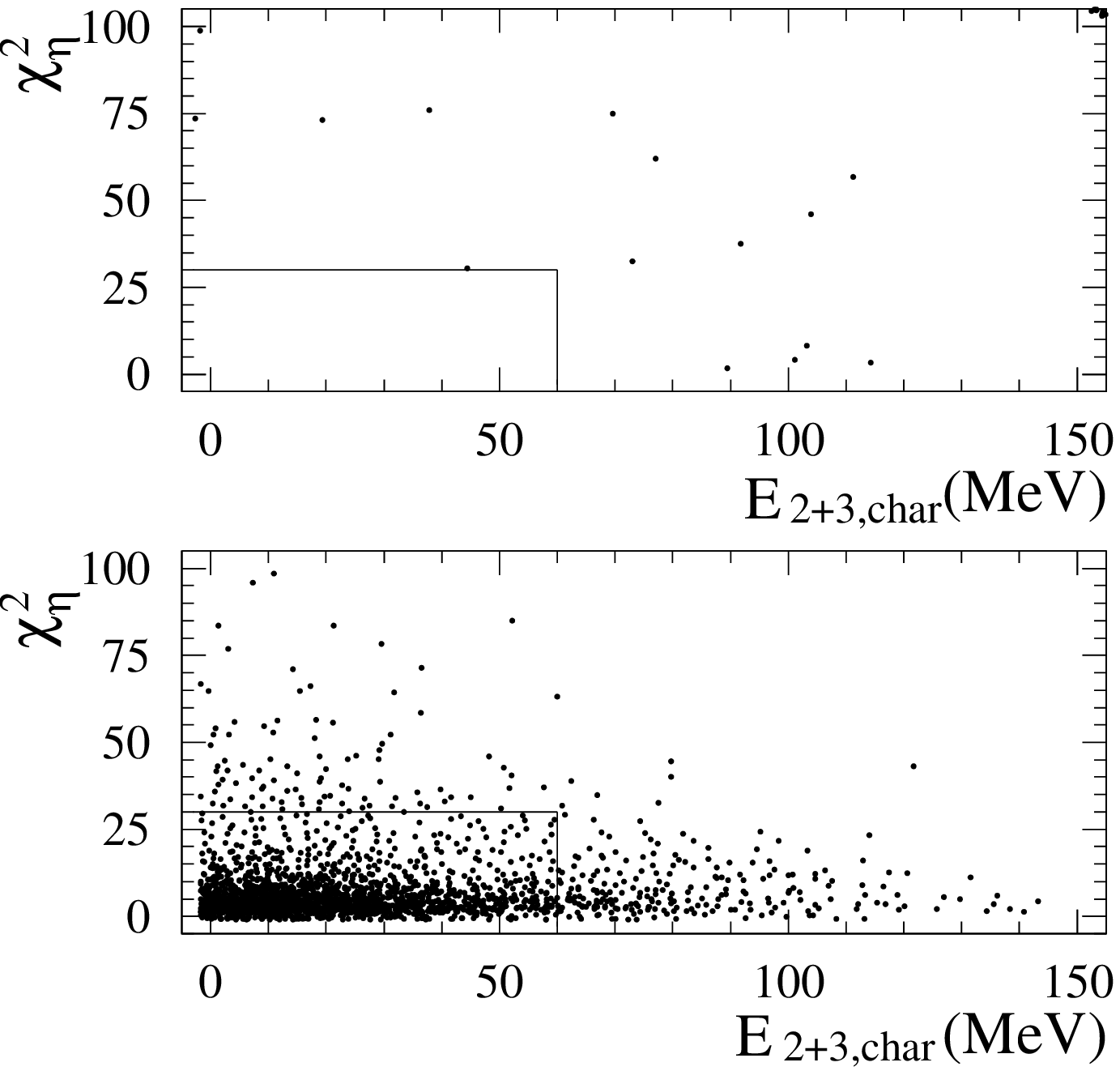}
\figcaption{\label{fig33} The two-dimensional distribution of 
$\chi^2_\eta$ versus $E_{\rm char,2+3}$ for data events (top) and 
simulated events of the 
$e^+e^-\to\eta^{\prime}\to\pi^+\pi^-\eta$, $\eta\to \gamma\gamma$ process
(bottom). The rectangle in the bottom left corner of the plot corresponds to
the selection criteria used.}
\end{center}

For events passing a preliminary selection  
the kinematic fit to the $e^+e^-\to\pi^+\pi^-\eta$ hypothesis 
is performed. The input parameters for kinematic fit are the polar and
azimuthal angles  of charged tracks and the angles and energies of photons
measured in the calorimeter. The quality of the fit is characterized by the 
parameter $\chi^2_\eta$.
Another important parameter used for the final selection is the
sum of energy depositions of charged particles in the second and third layers 
of the calorimeter $E_{\rm 2+3,char}$. Since pions in the process under study
are soft, they stop predominantly in the first calorimeter layer. 
The two-dimensional  distributions of the parameters $\chi^2_\eta$ and
$E_{\rm 2+3,char}$  for data events and simulated events of the
process under study are shown in Fig.~\ref{fig33}. The rectangle in the
bottom left corner corresponds to the selection criteria applied. 
No data events are selected.

The dominant sources of background for this decay mode are the processes 
$e^+e^-\to \eta\gamma, \eta\to \pi^+\pi^-\pi^0$ and
$e^+e^-\to \pi^+\pi^-\pi^0\pi^0$. Additional fake
photons can appear as a result of splitting of electromagnetic showers,
nuclear interaction of
pions in the calorimeter, or superimposing beam-generated background.
The number
of background events estimated using MC simulation
is $0.7\pm 0.1$ and
$0.10\pm 0.05$ for the first and second processes, respectively.

There is also the nonresonant reaction $e^+e^-\to \pi^+\pi^-\eta$, that 
proceeds
through the $\rho\eta$ intermediate state. It is 
suppressed due to the small phase space of the final particles.
The contribution of the nonresonant process is estimated to be 0.2 events.

\subsection{Decay chain $\eta^{\prime}\to\pi^+\pi^-\eta$, $\eta\to 3\pi^0$} 

For preliminary selected events the kinematic fit is performed to the
hypothesis $e^+e^-\to\pi^+\pi^-3\pi^0$.
The two-dimensional distributions of $\chi^2$ of the kinematic
fit ($\chi^2_{3\pi^0}$) versus the three $\pi^0$ invariant mass 
($M_{3\pi^0}$) for data events and simulated
events of $e^+e^-\to\eta^{\prime}\to\pi^+\pi^-\eta$, $\eta\to 3\pi^0$
process are shown in Fig.~\ref{fig4}.

\begin{center}
\includegraphics[width=6cm]{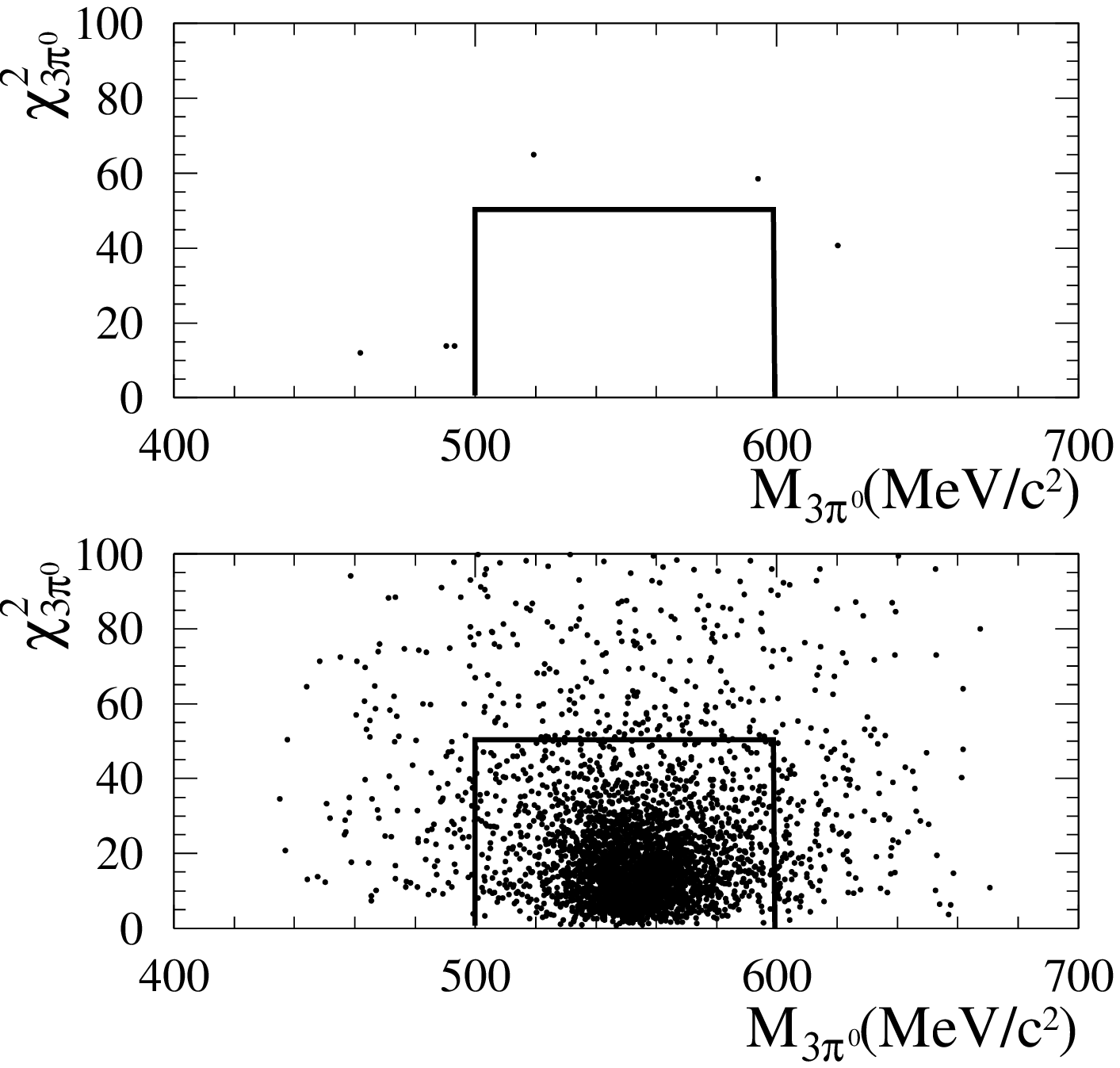}
\figcaption{\label{fig4} The two-dimensional distribution of the parameters $\chi^2_{3\pi^0}$
and $M_{3\pi^0}$ for data events (top) and simulated  
$\eta^{\prime}\to\pi^+\pi^-\eta, \eta\to 3\pi^0$ events (bottom).
The rectangle corresponds to the selection criteria used:
$\chi^2_{3\pi^0}<50$ and $500<M_{3\pi^0}<600$ MeV/$c^2$.}
\end{center}

The dominant background source for the $\pi^+\pi^-\pi^0\pi^0\pi^0$
final state is the process $e^+e^-\to \pi^+\pi^-\pi^0\pi^0$.
 The number of background events 
obtained using MC simulation is $2.7\pm 0.5$. 
The contribution of the nonresonant background from the 
$e^+e^-\to \pi^+\pi^-\eta$ process
discussed above is about 0.1 events.

\subsection{Decay chain $\eta^{\prime}\to\pi^0\pi^0\eta$, $\eta\to \gamma\gamma$ }

For events passing initial selection the kinematic fit to the 
$e^+e^-\to\eta^{\prime}\to\eta\pi^0\pi^0\to 6\gamma$ hypothesis is
performed. The quality of the fit is characterized by 
the parameter $\chi^2_{\eta\pi^0\pi^0}$. The distributions of this parameter
for data events, simulated signal events, and simulated background events 
from the process $e^+e^-\to\eta\gamma, \eta\to 3\pi^0$  are shown
in Fig.~\ref{fig6}.
The condition $\chi^2_{\eta\pi^0\pi^0} < 15$ is applied.
No data events satisfying the selection criteria have been found.

The main background sources for this decay mode 
are the processes $e^+e^-\to\eta\gamma\to 3\pi^0\gamma$ and
$e^+e^-\to \pi^0\pi^0\gamma$. 
The number of background events from these sources is 
calculated to be $1.3\pm 0.3$ and $0.4\pm 0.1$, respectively.

\subsection{Decay chain $\eta^{\prime}\to\pi^0\pi^0\eta$, $\eta\to 3\pi^0$ }
For this decay mode with ten photons in the final state there is no
background from $e^+e^-$ annihilation. The main source of background is
cosmic-ray showers. We select events containing nine or more photons
and no tracks in the drift chamber. 
The total energy 
deposition $E_{\rm cal}$ and the event momentum $P_{\rm cal}$ calculated 
using energy depositions in the calorimeter crystals must satisfy the 
following conditions:
\begin{eqnarray}
0.7 < E_{\rm cal}/E_{\rm cm} < 1.2,~cP_{\rm cal}/E_{\rm cm} < 0.3,\nonumber\\
~E_{\rm cal}/E_{\rm cm} - cP_{\rm cal}/E_{\rm cm} > 0.7.
\label{eton_vs_ptrt}
\end{eqnarray}
No data events are selected after applying these
criteria.

\begin{center}
\includegraphics[width=6cm]{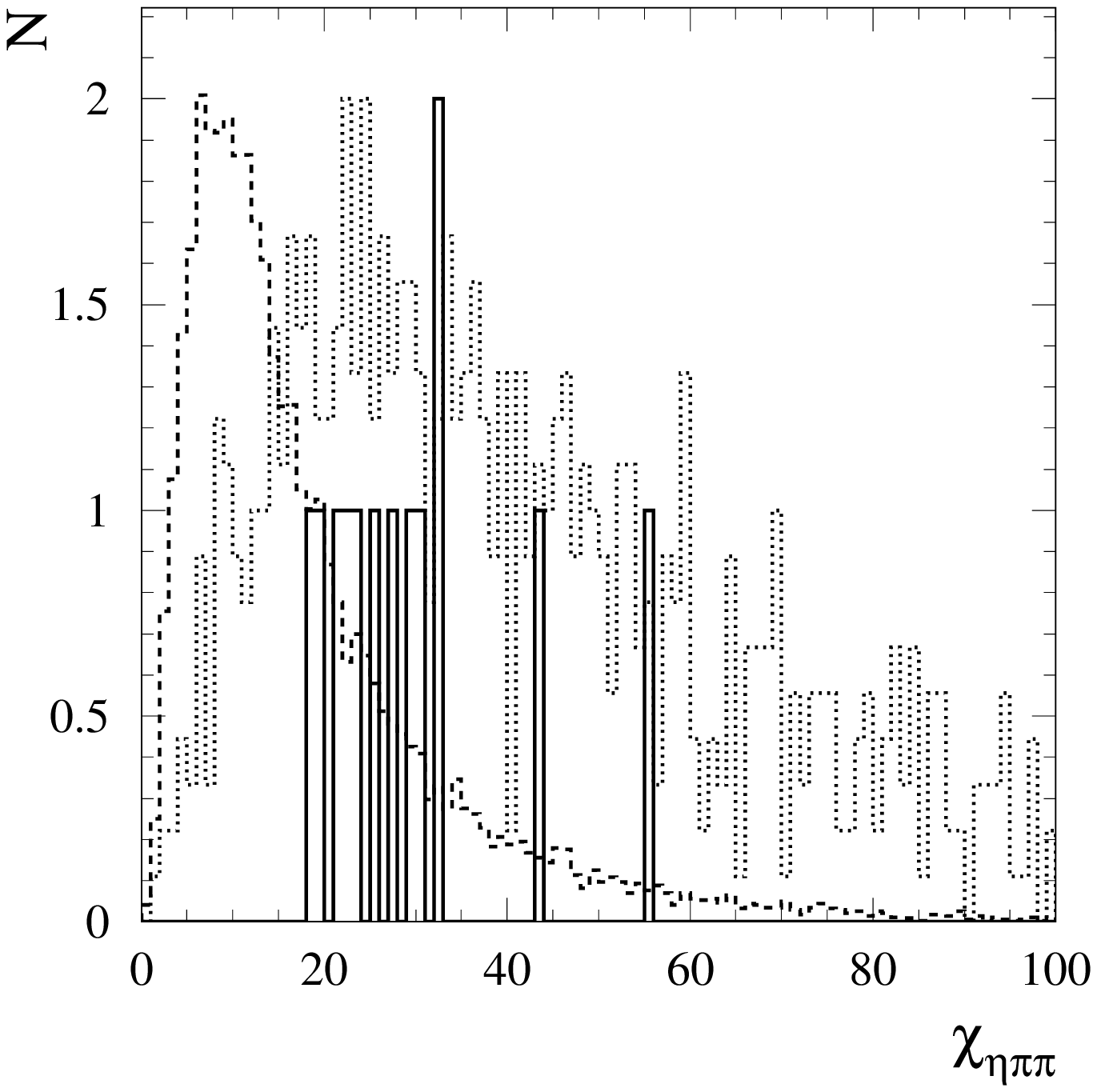}
\figcaption{\label{fig6} The $\chi^2_{\eta\pi^0\pi^0}$ distribution
for data events (solid histogram), simulated signal
$e^+e^-\to\eta^\prime \to 2\pi^0\eta \to 6\gamma$ events
(dashed histogram), simulated background events from the process 
$e^+e^-\to\eta\gamma, \eta\to 3\pi^0$ (dotted histogram).}
\end{center}

\subsection{Upper limit \label{ull}}
Since the number of selected data events is equal to zero, 
we set the upper limit on the cross section. 
The technique of Cousins and Highland~\cite{Cousins} following 
the implementation of Barlow~\cite{Barlow} is used to calculate the  
limit with all uncertainties included: 
\begin{equation}
\sigma_{\rm vis}^{\rm exp}<12.7\mbox{ pb at 90\% CL.}
\end{equation}
The limit on the cross section is translated using Eq.(\ref{xsvis}) to
the upper limit on the $\eta^{\prime}$ electronic width  
\begin{equation}
\Gamma_{\eta^{\prime}\to e^+e^-}< 0.0020\mbox{eV at 90\% CL.} 
\end{equation}

The obtained limit is slightly better than the limit set recently 
in the CMD-3 experiment $\Gamma_{\eta^{\prime}\to e^+e^-}< 0.0024$
eV~\cite{etapr-cmd}. 

Using the formula (\ref{sig1}) we
combine the SND and CMD-3 data and obtain the  
combined upper limits on the electronic width
\begin{equation}
\Gamma_{\eta^{\prime}\to e^+e^-}< 0.0011\mbox{ eV at 90\% CL}. 
\end{equation}
and the branching fraction
[$\Gamma_{\eta^{\prime}}=(0.198\pm0.009)$ MeV~\cite{pdg}]
\begin{equation}
{\cal B}(\eta^{\prime}\to e^+e^-) < 5.6\times10^{-9}
\mbox{ at 90\% CL}.
\end{equation}
The obtained upper limit is most stringent but still
30-50 times larger than theoretical
predictions~\cite{NPh1,NPh2} made in the framework of the Standard Model.

\section{Search for $\eta\to e^+e^-$ decay}

For this study, VEPP-2000 parameters at
c.m. energy close to $m_{\eta}c^2=548.862 \pm 0.018$ MeV~\cite{pdg}
such as luminosity, accuracy of the energy setting, energy
spread, are important. In 2013 SND did not record
data exactly at this energy. Therefore, we analyze data
from four energy points near $m_{\eta}c^2$, with c.m. energies of
520, 540, 560, and 580 MeV. The integrated luminosity
collected at these energy points measured using the
reaction $e^+e^-\to\gamma\gamma$ is $108.1 \pm 2.0$ nb$^{-1}$.

In the proposed experiment the collider energy must
be set and monitored with an accuracy better than
the collider c.m. energy spread of about 150 keV.
This is provided by the beam-energy-measurement system described above.

The most suitable $\eta$ decay mode for the search for
the $e^+e^-\to\eta$ reaction at SND is $\eta\to\pi^0\pi^0\pi^0\to 6\gamma$,
for which physical background is small. The main source
of background is cosmic-ray events. For the search for $e^+e^-\to\eta$,
events with six or more detected photons
and with the energy deposition in the calorimeter larger
than 0.6$E$ are selected. Background from events with
charged particles is rejected by the selection condition
that the number of fired wires in the drift chamber is
less than four. Cosmic-ray background is suppressed by
the veto from the muon detector.

For events passing the preliminary selection, a kinematic
fit to the $e^+e^-\to\pi^0\pi^0\pi^0\to 6\gamma$
hypothesis is performed. The
quality of the kinematic fit is characterized by the parameter $\chi^2$. 
The condition
$\chi^2<100$ is used to select
$\eta$ candidates. No events satisfying the selection criteria
are found. So, we set the upper limit on the $e^+e^-\to\eta$ cross section 
\begin{equation}
\sigma_{exp}<170 \mbox{ pb at 90\% CL.}
\end{equation}
corresponding to $N_s = 0$ and integrated luminosity 
108 nb$^{-1}$.
Using the same technique as in Sec.{\ref{ull}},
we can estimate sensitivity to the search for the decay $\eta\to e^+e^-$ to be
\begin{equation}
B(\eta\to e^+e^-)<2.9\times10^{-6} \mbox{ at 90\% CL.}
\end{equation}
This result is close to the upper limit $B(\eta\to e^+e^-)<2.3\times10^{-6}$
set recently in the HADES
experiment~\cite{hades}.  With a VEPP-2000 luminosity of $0.34\times10^{30}cm^{-2}sec^{-1}$
the current upper limit can
be reached in a week of data taking. In two weeks a
sensitivity at the level of $10^{-6}$
can be reached.

\section{Conclusion}

A search for the rare decay $\eta^{\prime} \to e^+e^-$ 
has been performed with the SND detector at the VEPP-2000 $e^+e^-$ collider. 
The inverse reaction
$e^+e^- \to \eta^{\prime}$ and five decay chains of $\eta^{\prime}$ have been
used for this search. 
The following upper limit has been set on the decay width: $\Gamma_{\eta^{\prime} \to e^+e^-}<0.002$ eV
at the 90\% confidence level. Also a sensitivity of SND in a search for decay $\eta \to e^+e^-$ has 
been studied. For this perpose we have analyzed a data sample with an integrated luminosity of 
108 nb$^{-1}$ collected with the SND detector in 
the center-of-mass energy range 520-580 MeV. There are no background events for the reaction  
$e^+e^- \to \eta$ with decay $\eta \to \pi^0\pi^0\pi^0$ have been found.
In the absence of background, a sensitivity to $B(\eta \to e^+e^-)$ of 10$^-6$ can be reached during 
two weeks of VEPP-2000 operation. 

\section{Acknowledgements}
Part of this work related to the photon
reconstruction algorithm in the electromagnetic calorimeter and 
analysis of multiphoton events is supported by Russian Science Foundation 
(project N 14-50-00080).
This work is partially supported by RFBR grant 
No.  15-02-03391.

\end{multicols}

\clearpage

\end{document}